\documentclass[%
 reprint,
 amsmath,amssymb,
 aps,
]{revtex4-2}
\usepackage{graphicx}
\usepackage{dcolumn}
\usepackage{bm}
\usepackage{amsthm}

\usepackage[colorlinks=true, allcolors=blue]{hyperref}
\usepackage{graphicx} 
\usepackage{braket}
\usepackage{amsfonts}
\usepackage{amsmath}
\usepackage{dsfont}
\usepackage{amsmath,amssymb}
\usepackage{tikz-cd}
\usepackage{quiver}
\usepackage{physics}
\usepackage{qcircuit}
\usepackage{dsfont}

\usepackage{xcolor}
\usepackage{soul}

\usepackage{algorithm}
\usepackage{algpseudocode}
\usepackage{mathtools}
\usepackage{lipsum}
\usepackage{float}
\usepackage[utf8]{inputenc}
\usepackage[justification=raggedright,singlelinecheck=false]{caption}

\usepackage{etoolbox}    

\usepackage{caption}
\usepackage{subfig}
\begin{document}

\preprint{APS/123-QED}

\title{Dissipative realization of a quantum distance-based classifier using open quantum walks}

\author{Pedro Linck Maciel}
\email{Contact author: pedro.linck@ufpe.br}
\affiliation{ Departamento de Física, Centro de Ciências Exatas e da Natureza, Universidade Federal de Pernambuco, Recife-PE, Brazil}

\author{Nadja K. Bernardes}
\affiliation{ Departamento de Física, Centro de Ciências Exatas e da Natureza, Universidade Federal de Pernambuco, Recife-PE, Brazil}

\author{Graeme Pleasance}
 \affiliation{School for Data Science and Computational Thinking, Stellenbosch University, Stellenbosch 7600, South Africa\\
National Institute for Theoretical and Computational Sciences (NITheCS), Stellenbosch 7600, South Africa}

\author{Francesco Petruccione}
 \affiliation{School for Data Science and Computational Thinking, Stellenbosch University, Stellenbosch 7600, South Africa\\
National Institute for Theoretical and Computational Sciences (NITheCS), Stellenbosch 7600, South Africa}



\date{\today}

\begin{abstract}
Open quantum walks (OQWs) constitute a class of quantum walks whose dynamics are entirely driven by interactions with the environment. It is well known that OQWs provide a general framework for implementing dissipative quantum computation. In this work, we demonstrate the feasibility of running the previously proposed quantum distance-based classifier within the open quantum walk computation model, and we show that its expected runtime remains finite even in the slower regime.
\end{abstract}

\maketitle

\section{\label{intro} Introduction}
\par Quantum computation, with its ability to explore the principles of quantum mechanics, has emerged as a promising paradigm for solving complex computational problems exponentially faster than classical computers. However, the development of practical and scalable quantum computers faces numerous challenges, including the detrimental effects of noise and decoherence. Quantum systems are inherently susceptible to environmental interactions, which can lead to loss of quantum information and degradation of computational performance.

\par To overcome these obstacles, researchers have been exploring a field known as dissipative quantum computation \cite{Cirac09, Petr12, Marshall2019}. Dissipative quantum computation aims to explore the unavoidable interactions between a quantum system and its surrounding environment, turning them from a detrimental effect into a resource for performing computational tasks. Instead of isolating the quantum system from its environment, as is commonly done in traditional quantum computing approaches, dissipative computation harnesses the dissipative dynamics to manipulate and process information.

\par An efficient implementation of dissipative quantum computing algorithms is the open quantum walk (OQW) \cite{Petr12(1), Petr12, Dutta25, Lardizabal2015Class}. Linear OQWs have been used for the design of algorithms \cite{Dutta25}, and its dynamics and computational limits have been established \cite{linck25}. OQWs offer a noise-resilient framework for dissipative quantum computation in which, at least in principle, one can engineer a reservoir so that the computation is driven entirely by the environment's dissipative dynamics. Moreover, for several OQW schemes there exists a useful class of convergence theorems \cite{Konno2013Limit,Attal2015CLT}. They also provide a natural way to incorporate a Markovian structure that is not readily available in standard coined quantum walks. In particular, linear open quantum walks have well-established convergence properties and a simple steady state $\cite{linck25}$. More broadly, because open system evolution extends beyond unitary dynamics, the OQW formalism allows one to explore general Kraus maps, opening the door to richer and more complex dynamical behavior.

\par Random walks, a fundamental concept in mathematics and physics, describe the stochastic motion of a walker in discrete steps, jumping from one node to another. Open quantum walks offer a novel framework for exploring the interplay between quantum dynamics and environmental noise, shedding light on the behavior of quantum systems under realistic and imperfect conditions. In this model, the transition between the nodes is driven purely by the dissipative interaction with an environment.

\par In the context of computation, a key area where quantum technology may prove to be beneficial is the field of machine learning \cite{Lloyd16, Schuld2019FeatureHilbert, Cong2019QCNN}. Quantum machine learning using open quantum systems has also demonstrated advantages \cite{Marshall2019DissipativeClassifier, korkmaz20, Brito2024OQSClassifier, Blank2020TailoredKernel, PerezSalinas2020DataReuploading}. Among the most basic machine-learning tasks is classification, typically implemented via classifier algorithms \cite{Hastie2009}. In this work, we focus on a linear classifier that has been simulated on a quantum computer by Schuld et. al. \cite{Schuld17, Neumann2021}. Recently, there have also been proposals exploring dissipative quantum systems to construct quantum classifiers \cite{Marshall2019, Turkpence2019, Zhang2021}. The idea is to encode the initialization information in the preparation of the physical system and the weight in the environmental factors (interaction and environmental states).

\par This work presents an implementation of a quantum distance-based classifier in an OQW model, as a proof of principle. Section~\ref{sec:preliminaries_oqw} summarizes the OQW results needed for computation in this framework. Section~\ref{sec:preliminaries_classifier} reviews the classifier and establishes, with proofs, both correctness and the expected post-selection success probability, extending earlier empirical claims. Section~\ref{sec:oqw_implementation} provides the OQW realization and validates the algorithm. Finally, a conclusion and future perspectives are presented in Section~\ref{sec:conclusion}.

\section{\label{sec:preliminaries_oqw}Open Quantum Walks } 
In this section, we develop the basic tools of open quantum walks needed to understand the results of this paper.

\subsection{Open Quantum Walks}
A quantum walk \cite{Venegas12, Ahar93} is a form of quantum evolution that occurs on a graph. The vertices of the graph form the basis for the graph space where the quantum walker is situated and can jump between nodes depending on both the graph structure and the internal walker state \cite{QWSA}. The dynamics, or time evolution, of a discrete quantum walk refers to the repeated application of the single-step evolution rule. Several quantum walks have been shown to model quantum computation \cite{Petr12, Childs09, Neil10}. Here, we focus on OQWs, a class of quantum walks whose dynamics is entirely induced by the environment.

\par A general quantum evolution (closed or open) is described as a map $\Lambda\colon \mathcal{B}(\mathcal{V}) \to \mathcal{B}(\mathcal{V})$, where $\mathcal{B}(\mathcal{V})$ denotes the algebra of bounded operators in the Hilbert space $\mathcal{V}$. The map $\Lambda$ must be linear, completely positive and trace-preserving \cite{OQS}. It is known \cite{OQS} that if $\text{dim}(\mathcal{V}) = d$, then we can write 
\begin{equation}
    \Lambda(\rho) = \sum_{i=0}^{k-1} K_i\rho K_i^\dagger,
\end{equation}
where $k \leq d^2$ and
\begin{equation}
    \sum_i K_i^\dagger K_i = \mathds{1}.
\end{equation}
This set of operators $\{K_i\}_{i=0}^{k-1}$ is called a set of Kraus operators for the map $\Lambda$.

\begin{figure}[t!]
\[\begin{tikzcd}
	& \cdots && \cdots \\
	\cdots & {\ket{i}} && {\ket{j}} & \cdots \\
	& \cdots && \cdots
	\arrow[curve={height=-6pt}, from=1-2, to=2-2]
	\arrow[curve={height=-6pt}, from=1-4, to=2-4]
	\arrow[curve={height=-6pt}, from=2-1, to=2-2]
	\arrow[curve={height=-6pt}, from=2-2, to=1-2]
	\arrow[curve={height=-6pt}, from=2-2, to=2-1]
	\arrow["{B_i^j}"{description}, curve={height=-12pt}, from=2-2, to=2-4]
	\arrow[curve={height=-6pt}, from=2-2, to=3-2]
	\arrow[curve={height=-6pt}, from=2-4, to=1-4]
	\arrow["{B_j^i}"{description}, curve={height=-12pt}, from=2-4, to=2-2]
	\arrow[curve={height=-6pt}, from=2-4, to=2-5]
	\arrow[curve={height=-6pt}, from=2-4, to=3-4]
	\arrow[curve={height=-6pt}, from=2-5, to=2-4]
	\arrow[curve={height=-6pt}, from=3-2, to=2-2]
	\arrow[curve={height=-6pt}, from=3-4, to=2-4]
\end{tikzcd}\]\caption{\label{fig:OQW} An arbitrary open quantum walk can be represented by this visual diagram. If there is an omitted edge in a particular diagram, this means that the corresponding operator $B_i^j$ is zero.}
\end{figure}
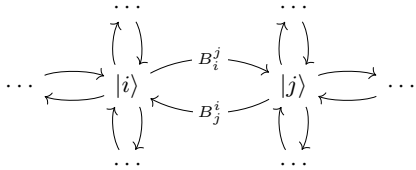

\par To formally define an open quantum walk, we need to define $\mathcal{H}$, the (internal) Hilbert space of the walker, and the graph Hilbert space $\mathcal{G}$ where the walk is made. We mark a basis $\{\ket{i}\}_{i \in G}$ for the graph space, where $G$ is the set of vertices of the underlying graph. For each $i \in G$ we also have a set $\{B_i^j\}_{j \in G}$ of linear operators $B_i^j \colon \mathcal{H}\to \mathcal{H}$ that satisfy
\begin{equation}
\label{eqn:kraus_cond_oqw}
    \sum_j B_i^{j \dagger} B_i^j = \mathds{1},
\end{equation}
which will act on the internal state when a jump is made from node $i$ to node $j$. In order to capture the jump together with this change in internal state, we define operators
\begin{equation}
M_i^j = B_i^j \otimes \ket{j}\bra{i} \in \mathcal{B}(\mathcal{H} \otimes \mathcal{G}).
\end{equation}
One can verify that Eq.~\ref{eqn:kraus_cond_oqw} implies
\begin{equation}
    \sum_{i,j} M_i^{j \dagger}M_i^j = \mathds{1},
\end{equation}
and therefore this set of operators defines a quantum channel. This set of Kraus operators defines an open quantum walk. An open quantum walk can be seen pictorially as a graph with labeled edges, where each edge from $i$ to $j$ is labeled $B_i^j$. Therefore, all information about the open quantum walk evolution map can be extracted from a given graphic representation. The recursive relations for an OQW with initial state $\rho_0 \in \mathcal{B}(\mathcal{H}\otimes \mathcal{G})$ are
\begin{equation}
\label{eqn:recursive_oqw}
    \begin{cases}
        \rho^{[0]} = \rho_0, \\
        \rho^{[n]} = \sum_{i,j}M_i^j\rho^{[n-1]}M_i^{j \dagger}, \; n\geq 1
    \end{cases}
\end{equation}
where $\rho^{[n]}$ is the state of the open quantum walk after $n$ steps. It is known \cite{Petr12(1)} that if $\rho_0 = \sum_{i,j} \rho_{ij}\otimes \ket{j}\bra{i}$ for some trace-class matrices $\rho_{ij}$ satisfying $\sum_i\text{Tr}(\rho_{ii}) = 1$, then after any $n\geq 1$ steps, $\rho^{[n]}$ has form 
\begin{equation}
    \rho^{[n]} = \sum_i \rho_{ii}^{[n]}\otimes \ket{i}\bra{i},
\end{equation}
where $\rho_{ii}^{[n]}$ are some trace-class matrices determined by the evolution defined by Eq.~\ref{eqn:recursive_oqw}. Note that if even a single step is made, the walker decouples from the graph space.

\begin{figure}[t!]
\[\begin{tikzcd}
	{\ket{0}} && {\ket{1}}
	\arrow["{B_0^0}", from=1-1, to=1-1, loop, in=60, out=120, distance=5mm]
	\arrow["{B_0^1}", curve={height=-6pt}, from=1-1, to=1-3]
	\arrow["{B_1^0}", curve={height=-6pt}, from=1-3, to=1-1]
	\arrow["{B_1^1}", from=1-3, to=1-3, loop, in=60, out=120, distance=5mm]
\end{tikzcd}\]\caption{\label{fig:oqw_example} An arbitrary two-node open quantum walk.}
\end{figure}
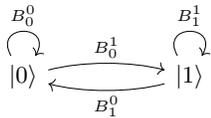

For example, for the two-node graph in Fig.~\ref{fig:oqw_example}, we have operators $B_0^0,B_0^1,B_1^0,B_1^1$ satisfying 
\begin{equation}
B_0^{0 \dagger}B_0^0 + B_0^{1 \dagger}B_0^1 = I,
\end{equation}
\begin{equation}
B_1^{1 \dagger}B_1^1 + B_1^{0 \dagger}B_1^0 = I.
\end{equation}
The state of the walker after $n$ steps will be
\begin{equation}
\rho^{[n]} = \rho_0^{[n]}\otimes \ket{0}\bra{0} + \rho_1^{[n]} \otimes \ket{1}\bra{1},
\end{equation}
where 
\begin{equation}
    \rho^{[n]}_0 = B_0^0 \rho_0^{[n-1]}B_0^{0 \dagger} + B_1^0\rho_1^{[n-1]}B_1^{0 \dagger},
\end{equation}
\begin{equation}
    \rho^{[n]}_1 = B_0^1 \rho_0^{[n-1]}B_0^{1 \dagger} + B_1^1\rho_1^{[n-1]}B_1^{1 \dagger}.
\end{equation}
If we have, for instance, $B_0^0 = \sqrt{\lambda}I, B_0^1 = \sqrt{\omega}U, B_1^0 = \sqrt{\lambda} U^\dagger, B_1^1 = \sqrt{\omega}I$, where $U$ is any unitary operator, $\omega, \lambda \geq 0$ are any non-negative real numbers satisfying $\omega + \lambda = 1$ and the initial state of the OQW is $\rho_0 = \ket{\psi}\bra{\psi} \otimes \ket{0}\bra{0}$, then we have \cite{linck25}
\begin{equation}
    \rho^{[n]} = \lambda \ket{\psi}\bra{\psi} \otimes \ket{0}\bra{0} + \omega U\ket{\psi}\bra{\psi}U^\dagger \otimes \ket{1}\bra{1}.
\end{equation}
In this way, if we want to perform quantum computation and apply a gate $U$ to the state $\ket{\psi}$, we have to post-select the result in node $\ket{1}$, which has success probability of $\omega$.

\subsection{Linear Open Quantum Walks for Quantum Computation}
\begin{figure}[b]
\[\begin{tikzcd}
	{\ket{0}} && {\ket{1}} && \cdots && {\ket{N-1}}
	\arrow["{\sqrt{\lambda} I}", from=1-1, to=1-1, loop, in=145, out=215, distance=10mm]
	\arrow["{\sqrt{\omega} U_0}", curve={height=-12pt}, from=1-1, to=1-3]
	\arrow["{\sqrt{\lambda} U_0^\dagger}", curve={height=-12pt}, from=1-3, to=1-1]
	\arrow["{\sqrt{\omega} U_1}", curve={height=-12pt}, from=1-3, to=1-5]
	\arrow["{\sqrt{\lambda} U_1^\dagger}", curve={height=-12pt}, from=1-5, to=1-3]
	\arrow["{\sqrt{\omega} U_{N-2}}", curve={height=-12pt}, from=1-5, to=1-7]
	\arrow["{\sqrt{\lambda} U_{N-2}^\dagger}", curve={height=-12pt}, from=1-7, to=1-5]
	\arrow["{\sqrt{\omega} I}"', from=1-7, to=1-7, loop, in=35, out=325, distance=10mm]
\end{tikzcd}\]\caption{\label{fig:linearOQW} The diagram corresponding to the linear OQW model. Each $U_i$ is a unitary operator, and $\omega, \lambda \geq 0$ are such that $\omega + \lambda = 1$.}
\end{figure}
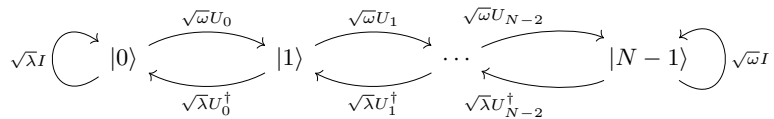

In this work, we use a linear topology of open quantum walk \cite{linck25} to perform quantum computation (i.e., the preparation of quantum states), where all Kraus operators of the evolution are proportional to unitaries. The graphic representation for this OQW model is shown in Fig.~\ref{fig:linearOQW}. Note that if $\rho_0 = \ket{\psi}\bra{\psi}\otimes \ket{0}\bra{0}$ we have (using the convention $U_{-1} = \mathds{1})$
\begin{equation}
\label{eqn:niteration}
\rho^{[n]} = \sum_i p_{i}^{(n)} U_{i-1}\cdots U_0\ket{\psi}\bra{\psi}U_0^\dagger \cdots U_{i-1}^\dagger \otimes \ket{i}\bra{i},
\end{equation}
where $p_i^{(n)}$ is the probability of finding the walker in node $i$ after $n$ steps. This can be seen as a classical Markov chain \cite{linck25} with a transition matrix \cite{MC,SP,HSM}
\begin{equation}
T=\left[\begin{array}{cccccc}
\lambda & \lambda & 0 & & & \\
\omega & 0 & \lambda & &  & \vdots \\
0 & \omega & 0 & & & \vdots \\
\vdots & 0 & \omega & & & \vdots \\
\vdots & & 0 & \ddots & & \vdots \\
 &  &  & & \ddots &  \\
 &  & 
\end{array}\right],
\end{equation}
and the steady-state $\pi$ can be calculated:
\begin{equation}
\pi_m = \dfrac{a^m(a-1)}{a^N-1},
\end{equation}
where $a = \dfrac{\omega}{\lambda} = \dfrac{\omega}{1-\omega}$.

\section{\label{sec:preliminaries_classifier}Distance-based classifier algorithms}
In this section, we develop the basic tools for classical and quantum distance-based classifiers needed to understand the results of this paper, and we provide rigorous proofs of several claims about the quantum distance-based classifier that were previously supported empirically.

\subsection{Classical distance-based classifier}
\label{subsec:classical_classifier}
We begin by highlighting the basic principles of a classical distance-based classifier~\cite{Hastie2009}. In a general classification task, one considers a vector space of parameters $\chi$ together with a finite set of possible classes $C = {0, 1, \dots, c-1}$, where each vector of parameters is uniquely associated with a class. As an illustrative example, the parameter vector may represent measurable attributes such as the height and weight of individuals, while the classes correspond to distinct categories, for instance, children and adults. The goal of the classification task is to infer the class associated with a given parameter vector. When a labeled dataset is available (meaning that the class corresponding to each parameter vector is known) the classification procedure is referred to as supervised learning.

Here, we focus on distance-based classifiers. This class of algorithms requires a kernel function $K\colon \chi \times \chi \to \mathbb{R}_{\ge 0}$, which is typically defined as an inner product $K(\vec{x},\vec{x}') = \langle \varphi(\vec{x}),\varphi(\vec{x}')\rangle$, where $\varphi\colon \chi \to \chi$ is a feature-mapping function and $\langle \; , \;\rangle$ can be any inner product, and in this work we use the usual inner product in $\mathbb{R}^N$: $\langle (x_1,\dots,x_N), (x_1', \dots ,x_N') \rangle = x_1x_1'  + \cdots + x_Nx_N'$. We assume $C = \{-1, 1\}$ and that we receive a dataset $D = \{(\vec{x}_m, y_m)\}_{m = 0}^{M-1}$, where each $\vec{x}_i$ is a parameter vector and $y_i \in C$ denotes its corresponding class. The inference rule is then defined as
\begin{equation}
    y = \text{sgn}\left(\sum_m y_m K(\vec{x}_m,\vec{x}) \right).
\end{equation}
This equation admits a natural probabilistic interpretation. Interpreting $K(\vec{x}_m, \vec{x})$ as an unnormalized probability weight, the inference rule assigns the class of the input parameter $\vec{x}$ according to the sign of its weighted mean value: a positive mean corresponds to $y = 1$, while a negative mean corresponds to $y = -1$. This interpretation is derived from the fact that when $\vec{x}$ is close to $\vec{x}_m$, their inner product approaches its maximum value. This will correspond, in the probabilistic picture, to a likelihood close to one. In this sense, two adjacent parameter vectors are highly likely to belong to the same class and $K$ can thus be viewed as a likelihood function that quantifies their proximity.

In this work, we adopt the following kernel function:
\begin{equation} \label{eqn:kernel}
K(\vec{x}, \vec{x}') = 1 - \frac{1}{4M} |\vec{x} - \vec{x}'|^2.
\end{equation}

\subsection{Quantum distance-based quantum classifier}
\label{subsec:quantum_classifier}

The main idea of the distance-based quantum classifier is to encode the data points in the amplitudes of a quantum state and, by exploring interference effects, to be able to predict the class of this point \cite{Schuld17}. In the original work \cite{Schuld17}, the algorithm is built such that it gives us the same prediction as the classical distance-based algorithm with the kernel function given by Eq.~\ref{eqn:kernel}.
\par The canonical way to embed a normalized parameter vector $\vec{x} = (x_1,\dots,x_{N})$ in a quantum state is to use a $N$-dimensional Hilbert space and encode as follows:
\begin{equation}
    \ket{\psi_{\vec{x}}} = \sum_{i \geq 1} x_i\ket{i}.
\end{equation}
To perform the quantum implementation of the distance-based classifier outlined in \ref{subsec:classical_classifier}, we first start with a state
\begin{equation}
    \ket{\mathcal{D}} = \dfrac{1}{\sqrt{2M}}\sum_{m = 0}^{M-1} \ket{m} (\ket{0}\ket{\psi_{\vec{x}}}+\ket{1}\ket{\psi_{\vec{x}_m}})\ket{y_m},
\end{equation}
where $\ket{y_m}=\ket{0}$ if $y_m = -1$, and $\ket{y_m} = \ket{1}$ if $y_m = 1$. The first system works as a flag, the second one is an ancillary qubit, and the third one encodes the parameter vectors. We perform a Hadamard gate in the ancilla qubit, obtaining
\begin{equation}
    \dfrac{1}{2\sqrt{M}}\sum_{m = 0}^{M-1} \ket{m} (\ket{0}\ket{\psi_{\vec{x}+\vec{x}_m}}+\ket{1}\ket{\psi_{\vec{x}-\vec{x_m}}})\ket{y_m},
\end{equation}
where $\ket{\psi_{\vec{x} \pm \vec{x}_m}} = \ket{\psi_{\vec{x}}}\pm \ket{\psi_{\vec{x}_m}}$. Now we perform a post-selection on the state of the ancilla, selecting the state $\ket{0}$:
\begin{equation}
    \dfrac{1}{2\sqrt{p_{\text{acc}} M}} \sum_{m=0}^{M-1}\sum_{i=1}^{N} \ket{m} (x^i + x_m^i)\ket{i} \ket{y_m},
\end{equation}
where
\begin{equation}
p_{\text{acc}} = \dfrac{1}{4M}\sum_m \braket{\psi_{\vec{x} + \vec{x}_m}}{\psi_{\vec{x} + \vec{x}_m} } = \dfrac{1}{4M}\sum_m ||\vec{x} + \vec{x}_m||^2
\end{equation}
is the success probability for the post-selection. It is also experimentally known \cite{Schuld17} that if the data is standardized (i.e., if the data is distributed with mean value equal to zero and standard deviation equal to one), then $p_{\text{acc}} \approx 1/2$. We prove this fact analytically, considering the data as a standardized random variable:
\begin{equation}
\begin{split}
    \mathbb{E}(p_{\text{acc}}) &= \dfrac{1}{4M}\sum_m \mathbb{E}[||\vec{x} + \vec{x}_m||^2] \\
    & = \dfrac{1}{4M}\sum_m \mathbb{E}[2 + 2\langle \vec{x},\vec{x}_m\rangle] \\
    & = \dfrac{1}{4M}\sum_m (2 + 2\langle \vec{x}, \mathbb{E}[\vec{x}_m]\rangle) \\
    & = \dfrac{1}{4M}\sum_m 2 \\
    & = \dfrac{1}{2},
\end{split}
\end{equation}
where $\mathbb{E}(\vec{x}_m) = 0$ since the dataset is standardized.
We define
\begin{equation}
    p(y = -1) = \dfrac{1}{4Mp_{\text{acc}}} \sum_{m \colon \ket{y_m} = \ket{0}}||\vec{x} + \vec{x}_m||^2,
\end{equation}
which corresponds to the sum of the amplitudes of each state with $\ket{y_m} = \ket{0}$. We also note that since the parameters are normalized, we have the following.
\begin{equation}
    \begin{split}
    \dfrac{1}{4Mp_{\text{acc}}} \sum_{m}||\vec{x} + \vec{x}_m||^2 + \dfrac{1}{4Mp_{\text{acc}}} \sum_{m}||\vec{x} - \vec{x}_m||^2 \\
    = \dfrac{1}{4M p_{\text{acc}}} \sum_m (2||\vec{x}||^2 + 2||\vec{x}_m||^2) 
    = \dfrac{1}{p_{\text{acc}}}.
    \end{split}
\end{equation}
We can therefore write
\begin{equation}
    p(y = -1) = \dfrac{1}{p_{\text{acc}}} \left(1- \dfrac{1}{4M}\sum_{m \colon \ket{y_m} = \ket{0}}||\vec{x} - \vec{x}_m||^2\right),
\end{equation}
and then we have 
\begin{equation}
    p(y = 1) = \dfrac{1}{p_{\text{acc}}} \left(1- \dfrac{1}{4M}\sum_{m \colon \ket{y_m} = \ket{1}}||\vec{x} - \vec{x}_m||^2\right).
\end{equation}
Since \(p(y=1)=1-p(y=-1)\), we predict \(y=+1\) if \(p(y=1) > p(y=-1)\) and \(y=-1\) otherwise. It is shown \cite{Schuld17} that this algorithm produces the same output as the classical distance-based classifier outlined in Section~\ref{subsec:classical_classifier}. Here, we give a formal proof of the correctness of this algorithm. First of all, we have the following.
\begin{equation}
    \mathbb{E}[y] = -1\cdot p(y = -1) + 1\cdot p(y = 1) = p(y = 1) - p(y = -1),
\end{equation}
and also have
\begin{equation}
    \begin{split}
       & \sum_m y_m \left(1 - \dfrac{1}{4N}||\vec{x}_m - \vec{x}||^2\right) =\\ 
       & \sum_{m \colon y_m = -1} -\left(1 - \dfrac{1}{4N}||\vec{x}_m - \vec{x}||^2\right) 
        + \\
        &\sum_{m \colon y_m = 1} \left(1 - \dfrac{1}{4N}||\vec{x}_m - \vec{x}||^2\right)
         =\\
    & p_{\text{acc}}(p(y=1)-p(y=-1))
         = p_{\text{acc}}\cdot \mathbb{E}[y].
    \end{split}\label{eq30}
\end{equation}
Note that the sign of $\mathbb{E}[y]$ determines the outcome of the quantum distance-based classifier, while the sign on the left-hand side of Eq.~\ref{eq30} corresponds to the result of the classical distance-based classifier. From this relation, if $p(y = -1) > p(y = 1)$, both the classical and quantum classifiers predict $y = -1$. The same reasoning applies to the case $y = 1$, demonstrating that the classical and quantum classifiers implement the same classification rule.

\subsection{Quantum circuit for distance-based classifier}
For a dataset containing two training examples with opposite labels,
$\mathcal{D}' = \{(\vec{x}_0, -1), (\vec{x}_1, 1)\}$, drawn from an arbitrarily large underlying dataset, and a test point $\vec{\tilde{x}}$ specified by two features, it was shown in Ref.~\cite{Neumann2021} that the corresponding classifier can be implemented using a simple two-qubit quantum circuit. We now follow the construction of this circuit. The feature vectors are defined as
\begin{equation}
\ket{x_0} = \cos\left(\tfrac{\theta}{2}\right)\ket{0} - \sin\left(\tfrac{\theta}{2}\right)\ket{1},
\end{equation}
\begin{equation}
\ket{x_1} = \cos\left(\tfrac{\phi}{2}\right)\ket{0} - \sin\left(\tfrac{\phi}{2}\right)\ket{1},
\end{equation}
\begin{equation}
\ket{\tilde{x}} = \cos\left(\tfrac{\gamma}{2}\right)\ket{0} - \sin\left(\tfrac{\gamma}{2}\right)\ket{1},
\end{equation}
where $\theta$ can be set to zero without loss of generality, such that $\ket{x_0} = \ket{0}$. In this case, the initial state used to perform the distance-based quantum classification has identical class and flag registers; therefore, a single qubit suffices to encode both.
\begin{equation}
\frac{1}{2}\ket{0}\left(\ket{0}\ket{\tilde{x}} + \ket{1}\ket{x_0}\right)
+ \frac{1}{2}\ket{1}\left(\ket{0}\ket{\tilde{x}} + \ket{1}\ket{x_1}\right).
\end{equation}

The ratio $t$ between the probabilities of measuring $\ket{0}$ and $\ket{1}$ in the class register is given by
\begin{equation}
t = \frac{P(\ket{y} = \ket{0})}{P(\ket{y} = \ket{1})}
= \frac{\cos^2(\gamma/4)}{\cos^2\big((\gamma - \phi)/4\big)}.
\end{equation}

The circuit implementing the reduced quantum distance-based classifier with two qubits is shown below. The second register corresponds to the $\ket{y}$ (class) qubit, while the first acts as an ancilla.

\[ \Qcircuit @C=2em @R=1em {
& \lstick{\ket{0}} & \gate{H} & \ctrl{1} & \gate{H} & \qw \\
& \lstick{\ket{0}} & \gate{R_y(-\omega'/2)} & \gate{X} & \gate{R_y(\omega'/2)} & \qw\\
} \]

Here, $\omega'$ must be determined to reproduce the quantum algorithm described in Sec.~\ref {subsec:quantum_classifier}. The portion of the quantum circuit responsible for state preparation comprises all gates except the final Hadamard gate. The state-preparation stage of the circuit yields
\begin{equation}
\dfrac{1}{\sqrt{2}}(\cos(\omega'/2)\ket{11} - \sin(\omega'/2)\ket{10} + \ket{00}).
\end{equation}
After the last Hadamard gate, the final state $\ket{\psi}$ is then 
\begin{equation}
\begin{split}
\ket{\psi} & = \dfrac{1}{2}(-\cos(\omega'/2)\ket{11}   + \cos(\omega'/2)\ket{01} \\
        &+ (1 + \sin(\omega'/2))\ket{10} + (1-\sin(\omega'/2))\ket{00} ).
\end{split}
\end{equation}
After post-selecting $\ket{0}$ in the first register, we obtain the state $\ket{\psi'}$
\begin{equation}
    \ket{\psi'} = \dfrac{1}{2\sqrt{p'_{\text{acc}}}}(\cos(\omega'/2)\ket{01} + (1-\sin(\omega'/2))\ket{00}),
\end{equation}
where 
\begin{equation}
    p'_{\text{acc}} = \dfrac{ 1 - \sin(\omega'/2)}{2}.
\end{equation}
Classifying $y$ based on the second qubit register, we obtain
\begin{equation}
    \dfrac{P'(\ket{y} = \ket{0})}{P'(\ket{y} = \ket{1})} =  \dfrac{(1-\sin(\omega'/2))^2}{\cos^2(\omega'/2)}.
\end{equation}
To obtain the same distribution probability of the quantum distance-based classifier, we impose the fact that the above equation is also equal to $t$. This gives us 
\begin{equation}\label{eqn:trig1}
    \dfrac{(1-\sin(\omega'/2))^2}{\cos^2(\omega'/2)} = t,
\end{equation}
which, solving for $\omega'$ (see Appendix \ref{app:trigonometry}), gives us 
\begin{equation}\label{eqn:trig2}
    \omega ' = 4 \arctan \left( \dfrac{1-\sqrt{t}}{1+\sqrt{t}}\right).
\end{equation}

\section{\label{sec:oqw_implementation}Distance-based quantum classifier in the OQW model}
In this section, we implement the quantum distance-based classifier proposed by Schuld \cite{Schuld17, Neumann2021} in the open quantum walk model of quantum computation.

\subsection{Implementing the circuit in the framework}

We now show that the open quantum walk can be used to implement the quantum classifier described above.We restrict ourselves to the minimal two-feature implementation, since our goal is to provide a proof of principle that the algorithm can be feasibly implemented and executed within this framework. To this end, we consider the simplest possible scenario: two classes and two features. The minimal circuit executing this classifier is presented in Sec.~\ref{subsec:quantum_classifier} and consists of two qubits and three layers of gates. Within the OQW framework, this circuit is implemented through the following unitaries acting on the two qubits:\begin{eqnarray}
U_1=H\otimes R_y(-\omega’/2),\\
U_2=CNOT,\\
U_3=H\otimes R_y(\omega’/2).
\end{eqnarray}
and is represented by a graph with four nodes, as depicted in Fig.~\ref{fig:OQW}. Typically, each additional circuit layer is implemented in the OQW model by introducing one extra node together with its associated unitary, so that the graph size increases linearly with the circuit depth. Note that the internal degrees of freedom of the walker are represented by two qubits, and the graph itself is also encoded in two qubits; therefore, a total of four qubits are required for this implementation. The complete set of transition operators is given by $\{B_0^0=\sqrt{\lambda}I,B_0^1=\sqrt{\omega}U_1,B_1^0=\sqrt{\lambda}U_1^{\dagger},B_1^2=\sqrt{\omega}U_2,B_2^1=\sqrt{\lambda}U_2^{\dagger},B_2^3=\sqrt{\omega}U_3,B_3^2=\sqrt{\lambda}U_3^{\dagger},B_3^3=\sqrt{\omega}I\}$.

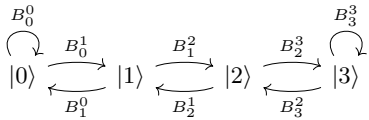
\begin{figure}
    \centering
\[\begin{tikzcd}
	{\ket{0}} & {\ket{1}} & {\ket{2}} & {\ket{3}}
	\arrow["{B_0^0}", from=1-1, to=1-1, loop, in=60, out=120, distance=5mm]
	\arrow["{B_0^1}", curve={height=-6pt}, from=1-1, to=1-2]
	\arrow["{B_1^0}", curve={height=-6pt}, from=1-2, to=1-1]
	\arrow["{B_1^2}", curve={height=-6pt}, from=1-2, to=1-3]
	\arrow["{B_2^1}", curve={height=-6pt}, from=1-3, to=1-2]
	\arrow["{B_2^3}", curve={height=-6pt}, from=1-3, to=1-4]
	\arrow["{B_3^2}", curve={height=-6pt}, from=1-4, to=1-3]
	\arrow["{B_3^3}", from=1-4, to=1-4, loop, in=60, out=120, distance=5mm]
\end{tikzcd}\]
    \caption{Open Quantum Walk representing the quantum classifier circuit.}
    \label{fig:OQW-circuit}
\end{figure}

\begin{figure}[H]
\centering
\subfloat[\centering]{\includegraphics[width=6.0cm]{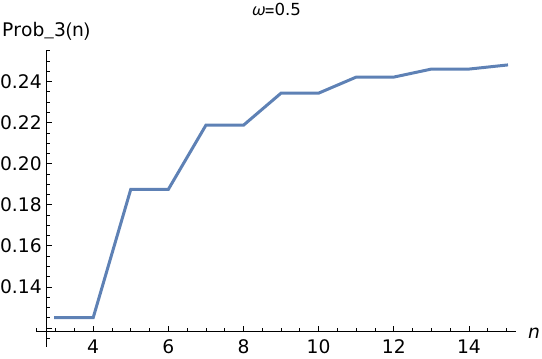}}\vspace{0.5cm}\\
\subfloat[\centering]{\includegraphics[width=6.0cm]{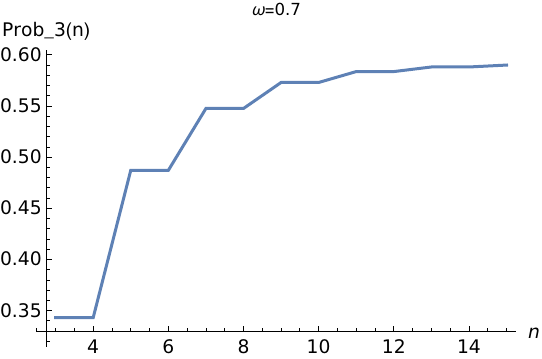}}\vspace{0.5cm}\\
\subfloat[\centering]{\includegraphics[width=6.0cm]{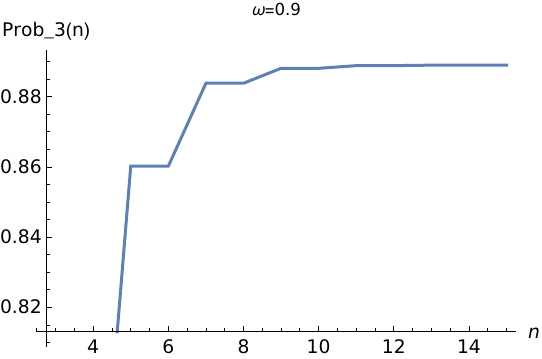}}
\caption{Detection probability of finding the walker in node 3 $P_3(n)$ versus the number of iterations $n$ for different values of probability $\omega$.\label{ProbOQW}}
\end{figure}

To determine how many steps are required for the OQW to reach steady state, we initialize the walker in the state $\ket{00}\bra{00}\otimes\ket{00}\bra{00}$. The first two qubits represent the system that enters the circuit, while the last two qubits encode the node, labeled from 0 to 3. Since the result of the computation is always found in the last node, we first evaluate the detection probability at node 3 as a function of the number of OQW iterations $n$ for different values of $\omega$, as shown in Fig.~\ref{ProbOQW}. From these results, we observe that 10 iterations are sufficient for the algorithm to converge. 

\begin{figure*}[t!]
\scalebox{0.58}{\input{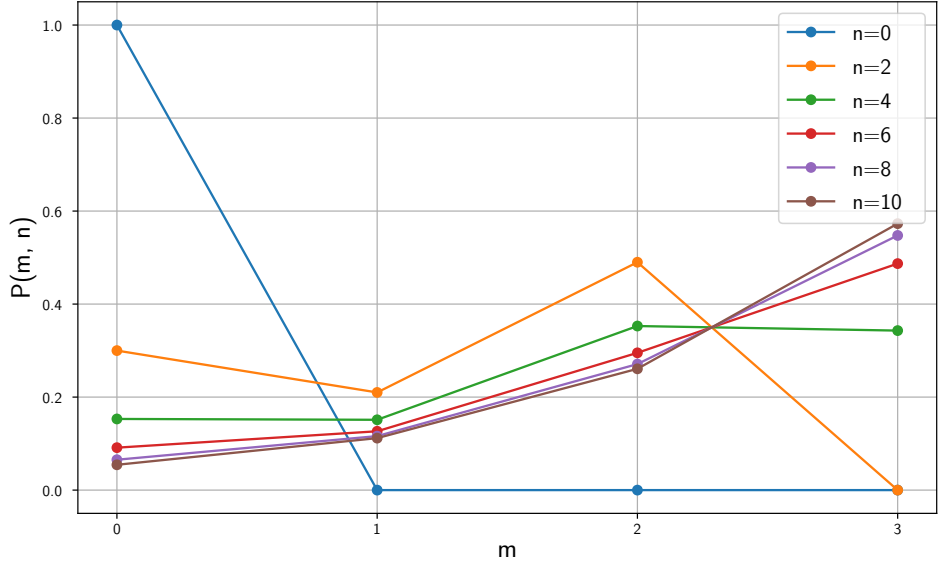}}
\caption{\label{fig:evolution} Evolution of probabilities in the linear OQW with $N=4$ and $\omega=0.7$. Distributions are shown at time steps $n=0,2,\dots,10$.}
\end{figure*}

Alternatively, one could employ the analytical estimate derived in Ref.~\cite{linck25} for $\omega > 1/2$ to obtain\begin{equation} \label{eqn:iteration_number}
    n_{\text{iterations}}(\omega) = \dfrac{N}{2\omega-1},
\end{equation}
which in our case reduces to 
\begin{equation}
    n_{\text{iterations}}(\omega = 0.7) = 10
\end{equation}
\begin{equation}
    n_{\text{iterations}}(\omega = 0.9) = 5.
\end{equation}
In what follows, we use 10 iterations for all three values of $\omega$ to approach the steady state. However, it should be noted that these additional steps increase the probability of detection at the last node by only about 2\% for $\omega = 0.9$. This improvement can be achieved with minimal computational cost as the underlying graph is small. In Fig.~\ref{ProbOQW}, the detection probability of finding the walker in the last node is shown for different values of $\omega$. In Fig.~\ref{fig:evolution}, we can see the evolution of probabilities for $\omega = 0.7$. Also note that, although we are using a graph of size $4$, Eq. $~\ref{eqn:iteration_number}$ is fully general and shows that the convergence time grows linearly with the graph size.

\subsection{Simulation results}

Now we can implement the classifier. We will consider the Iris dataset and will analyze only the setosa and versicolor classes and the width and length of the sepal feature. In this way, our data set will be formed by a 2-dimensional vector and will contain 100 vectors. We will work with a standardized and normalized data set.

To implement the circuit presented in the previous section, we essentially need to determine $\omega'$. To this end, we randomly select two vectors, each belonging to one of the two distinct classes, and a third vector corresponding to the unknown point. These three data points are represented by
\begin{eqnarray}
\ket{x_0}=\ket{0},\\
\ket{x_1}=\cos{(\phi/2)}\ket{0}+\sin{(\phi/2)}\ket{1},\\
\ket{\tilde{x}}=\cos{(\gamma/2)}\ket{0}+\sin{(\gamma/2)}\ket{1}.
\end{eqnarray}
For example, for Iris samples 34, 75, and 13, the standardized and normalized data vectors are given by $x_0 = (0.999807, 0.0196469)$, $x_1 = (-0.275974,, 0.961165)$, and $\tilde{x} = (-0.194006, -0.981000)$. The state $x_0$ is taken as the reference, and in the rotated reference frame it becomes $x_0 = (0, 1)$, $x_1 = (-0.966402, -0.257036)$, and $\tilde{x} = (0.976999, -0.213242)$. This corresponds to the rotation angles $\phi = -8.90487$ and $\gamma = -3.57138$, leading to $\omega' = 1.06501$.

We then ran the quantum classifier for 2000 triples of points, and the probability of correctly identifying the class of the unknown point is shown in Table \ref{tab:classifier_success_error} for different values of the parameter $\omega$, averaged over 10 iterations. In the table, it shows the probability of success $P_{\text{succ}}$ in obtaining the correct class, the probability $P_{\text{error}}(1|2)$ in predicting the class $1$ when the true class is $2$ and the probability $P_{\text{error}}(2|1)$ in predicting the class $2$ when the true class is $1$.

Note that the probability $P_{\text{succ}}$ is independent of the parameter $\omega$ because of the post-selection procedure. Nevertheless, $\omega$ still shapes the system's dynamics: it controls the convergence rate and, in the long-time regime, the population accumulated at the final node. Hence, although $\omega$ does not change the classification outcome, it does impact the runtime of the model by modulating how quickly the computation reaches the post-selected event.

\begin{table}[t] 
\caption{Success probability and conditional error rates of the classifier as a function of $\omega$. Here, $P_{\text{succ}}$ is the probability of predicting the correct class, $P_{\text{error}}(1|2)$ is the probability of predicting class 1 when the true class is 2, and $P_{\text{error}}(2|1)$ vice versa.\label{tab:classifier_success_error}}
\begin{ruledtabular}
\begin{tabular}{cccc}
$\omega$ & $P_{\text{succ}}$ & $P_{\text{error}}(1|2)$ & $P_{\text{error}}(2|1)$ \\
\colrule
1 & 91.3\% & 7.9\% & 0.8\% \\
0.8 & 89.5\% & 0.5\% & 1.0\% \\
0.5 & 90.4\% & 8.4\% & 1.2\% \\
\end{tabular}
\end{ruledtabular}
\end{table}

\section{\label{sec:conclusion}Conclusion}
We have demonstrated that the quantum distance-based classifier can be faithfully implemented within the open quantum walk (OQW) model for dissipative quantum computation and that its performance is independent of the environmental parameters $\omega$ and $\lambda$. By deriving analytical expressions for the classifier’s correctness and post-selection success probability, we provided a rigorous connection between the quantum and classical distance-based frameworks. These results show that OQWs can serve as a viable platform to realize basic quantum machine-learning algorithms in a dissipative setting.

\par Our simulations confirm that the OQW implementation is feasible and reproduces the expected classification behavior for two-feature data sets, such as the Iris benchmark, thereby providing the proof of principle we originally aimed to demonstrate. However, it is important to emphasize that we did not modify the algorithm; rather, we embedded it in the OQW model in order to demonstrate its feasibility. It is also important to note that although there are implementations of algorithms in the OQW framework, such as the Fourier transform \cite{Petr12} and the Page-Rank \cite{Dutta25}, the literature on such implementations is still scarce. We do not compare our results with other implementations of this algorithm, such as Schuld's $~\cite{Schuld17}$, because Schuld’s approach is circuit-based, whereas in our model the unitaries are implemented via interaction with the environment. A direct comparison is therefore not entirely fair, since the two settings correspond to different computational models.

\par Finally, although we have employed a restricted OQW model in which all Kraus operators are proportional to unitaries, extending this framework to include non-unitary Kraus operators may enable more efficient dissipative algorithms. These directions represent promising avenues for future theoretical and experimental research.

\begin{acknowledgments}
This study was financed in part by the Coordenação de Aperfeiçoamento de Pessoal de Nível Superior – Brazil (CAPES) – Finance Code 001. N.K.B. acknowledges financial support from CNPq Brazil (442429/2023-1) and FAPESP(Grant 2021/06035-0). The authors acknowledge the financial support of the National Institute of Science and Technology for Applied Quantum Computing through CNPq process No. 408884/2024-0.
\end{acknowledgments}

\appendix
\label{app:trigonometry}
\section{}
Here we show, for completeness, that Eq.~\ref{eqn:trig2} follows from Eq.~\ref{eqn:trig1}. 
We first prove that, for any $x\in\mathbb{R}$ for which the expressions are well defined,
\begin{equation}
    \frac{1-\tan(x/2)}{1+\tan(x/2)}=\frac{1-\sin x}{\cos x}.
\end{equation}
Indeed,
\begin{equation}
\begin{split}
\frac{1-\sin x}{\cos x}
&= \frac{1-\sin\!\bigl(2\cdot \frac{x}{2}\bigr)}{\cos\!\bigl(2\cdot \frac{x}{2}\bigr)} \\
&= \frac{1-2\sin\!\bigl(\frac{x}{2}\bigr)\cos\!\bigl(\frac{x}{2}\bigr)}
        {\cos^2\!\bigl(\frac{x}{2}\bigr)-\sin^2\!\bigl(\frac{x}{2}\bigr)} \\
&= \frac{\sec^2\!\bigl(\frac{x}{2}\bigr)-2\tan\!\bigl(\frac{x}{2}\bigr)}
        {1-\tan^2\!\bigl(\frac{x}{2}\bigr)} \\
&= \frac{1+\tan^2\!\bigl(\frac{x}{2}\bigr)-2\tan\!\bigl(\frac{x}{2}\bigr)}
        {1-\tan^2\!\bigl(\frac{x}{2}\bigr)} \\
&= \frac{\Bigl(1-\tan\!\bigl(\frac{x}{2}\bigr)\Bigr)^2}
        {1-\tan^2\!\bigl(\frac{x}{2}\bigr)} \\
&= \frac{1-\tan\!\bigl(\frac{x}{2}\bigr)}
        {1+\tan\!\bigl(\frac{x}{2}\bigr)} .
\end{split}
\end{equation}
The second equality uses the double-angle formulas for $\sin$ and $\cos$; the third is obtained by multiplying numerator and denominator by $\sec^2(x/2)$; the fourth uses $\sec^2(x/2)=1+\tan^2(x/2)$; and the remaining steps follow from straightforward algebra. Therefore, Eq.~\ref{eqn:trig1} implies
\begin{equation}
    \frac{1-\sin(\omega'/2)}{\cos(\omega'/2)}
    = \frac{1-\tan(\omega'/4)}{1+\tan(\omega'/4)}
    = \sqrt{t},
\end{equation}
from which we may isolate $\tan(\omega'/4)$ to obtain
\begin{equation}
    \tan(\omega'/4) = \frac{1-\sqrt{t}}{1+\sqrt{t}}.
\end{equation}
This is equivalent to Eq.~\ref{eqn:trig2} (up to the choice of branch and $2\pi$ periodicity).

\newpage

\nocite*
\bibliography{references}

\providecommand{\noopsort}[1]{}\providecommand{\singleletter}[1]{#1}%
\begin{thebibliography}{31}%
\makeatletter
\providecommand \@ifxundefined [1]{%
 \@ifx{#1\undefined}
}%
\providecommand \@ifnum [1]{%
 \ifnum #1\expandafter \@firstoftwo
 \else \expandafter \@secondoftwo
 \fi
}%
\providecommand \@ifx [1]{%
 \ifx #1\expandafter \@firstoftwo
 \else \expandafter \@secondoftwo
 \fi
}%
\providecommand \natexlab [1]{#1}%
\providecommand \enquote  [1]{``#1''}%
\providecommand \bibnamefont  [1]{#1}%
\providecommand \bibfnamefont [1]{#1}%
\providecommand \citenamefont [1]{#1}%
\providecommand \href@noop [0]{\@secondoftwo}%
\providecommand \href [0]{\begingroup \@sanitize@url \@href}%
\providecommand \@href[1]{\@@startlink{#1}\@@href}%
\providecommand \@@href[1]{\endgroup#1\@@endlink}%
\providecommand \@sanitize@url [0]{\catcode `\\12\catcode `\$12\catcode `\&12\catcode `\#12\catcode `\^12\catcode `\_12\catcode `\%12\relax}%
\providecommand \@@startlink[1]{}%
\providecommand \@@endlink[0]{}%
\providecommand \url  [0]{\begingroup\@sanitize@url \@url }%
\providecommand \@url [1]{\endgroup\@href {#1}{\urlprefix }}%
\providecommand \urlprefix  [0]{URL }%
\providecommand \Eprint [0]{\href }%
\providecommand \doibase [0]{https://doi.org/}%
\providecommand \selectlanguage [0]{\@gobble}%
\providecommand \bibinfo  [0]{\@secondoftwo}%
\providecommand \bibfield  [0]{\@secondoftwo}%
\providecommand \translation [1]{[#1]}%
\providecommand \BibitemOpen [0]{}%
\providecommand \bibitemStop [0]{}%
\providecommand \bibitemNoStop [0]{.\EOS\space}%
\providecommand \EOS [0]{\spacefactor3000\relax}%
\providecommand \BibitemShut  [1]{\csname bibitem#1\endcsname}%
\let\auto@bib@innerbib\@empty
\bibitem [{\citenamefont {Verstraete}\ \emph {et~al.}(2009)\citenamefont {Verstraete}, \citenamefont {Wolf},\ and\ \citenamefont {Cirac}}]{Cirac09}%
  \BibitemOpen
  \bibfield  {author} {\bibinfo {author} {\bibfnamefont {F.}~\bibnamefont {Verstraete}}, \bibinfo {author} {\bibfnamefont {M.~M.}\ \bibnamefont {Wolf}},\ and\ \bibinfo {author} {\bibfnamefont {J.~I.}\ \bibnamefont {Cirac}},\ }\bibfield  {title} {\bibinfo {title} {Quantum computation and quantum-state engineering driven by dissipation},\ }\href {https://doi.org/10.1038/nphys1342} {\bibfield  {journal} {\bibinfo  {journal} {Nature Physics}\ }\textbf {\bibinfo {volume} {5}},\ \bibinfo {pages} {633} (\bibinfo {year} {2009})}\BibitemShut {NoStop}%
\bibitem [{\citenamefont {Sinayskiy}\ and\ \citenamefont {Petruccione}(2012)}]{Petr12}%
  \BibitemOpen
  \bibfield  {author} {\bibinfo {author} {\bibfnamefont {I.}~\bibnamefont {Sinayskiy}}\ and\ \bibinfo {author} {\bibfnamefont {F.}~\bibnamefont {Petruccione}},\ }\bibfield  {title} {\bibinfo {title} {Efficiency of open quantum walk implementation of dissipative quantum computing algorithms},\ }\href {https://doi.org/10.1007/s11128-012-0426-3} {\bibfield  {journal} {\bibinfo  {journal} {Quantum Inf Process}\ }\textbf {\bibinfo {volume} {11}},\ \bibinfo {pages} {1301} (\bibinfo {year} {2012})}\BibitemShut {NoStop}%
\bibitem [{\citenamefont {Marshall}\ \emph {et~al.}(2019{\natexlab{a}})\citenamefont {Marshall}, \citenamefont {Venuti},\ and\ \citenamefont {Zanardi}}]{Marshall2019}%
  \BibitemOpen
  \bibfield  {author} {\bibinfo {author} {\bibfnamefont {J.}~\bibnamefont {Marshall}}, \bibinfo {author} {\bibfnamefont {L.~C.}\ \bibnamefont {Venuti}},\ and\ \bibinfo {author} {\bibfnamefont {P.}~\bibnamefont {Zanardi}},\ }\bibfield  {title} {\bibinfo {title} {Classifying quantum data by dissipation},\ }\href {https://doi.org/10.1103/PhysRevA.99.032330} {\bibfield  {journal} {\bibinfo  {journal} {Physical Review A}\ }\textbf {\bibinfo {volume} {99}},\ \bibinfo {pages} {032330} (\bibinfo {year} {2019}{\natexlab{a}})}\BibitemShut {NoStop}%
\bibitem [{\citenamefont {Attal}\ \emph {et~al.}(2012)\citenamefont {Attal}, \citenamefont {Petruccione}, \citenamefont {Sabot},\ and\ \citenamefont {Sinayskiy}}]{Petr12(1)}%
  \BibitemOpen
  \bibfield  {author} {\bibinfo {author} {\bibfnamefont {S.}~\bibnamefont {Attal}}, \bibinfo {author} {\bibfnamefont {F.}~\bibnamefont {Petruccione}}, \bibinfo {author} {\bibfnamefont {C.}~\bibnamefont {Sabot}},\ and\ \bibinfo {author} {\bibfnamefont {I.}~\bibnamefont {Sinayskiy}},\ }\bibfield  {title} {\bibinfo {title} {Open quantum random walks},\ }\href {https://doi.org/10.1007/s10955-012-0491-0} {\bibfield  {journal} {\bibinfo  {journal} {J Stat Phys}\ }\textbf {\bibinfo {volume} {147}},\ \bibinfo {pages} {832–} (\bibinfo {year} {2012})}\BibitemShut {NoStop}%
\bibitem [{\citenamefont {Dutta}(2025)}]{Dutta25}%
  \BibitemOpen
  \bibfield  {author} {\bibinfo {author} {\bibfnamefont {S.}~\bibnamefont {Dutta}},\ }\bibfield  {title} {\bibinfo {title} {Discrete-time open quantum walks for vertex ranking in graphs},\ }\href {https://doi.org/https://doi.org/10.1103/PhysRevE.111.034312} {\bibfield  {journal} {\bibinfo  {journal} {Phys. Rev. E}\ }\textbf {\bibinfo {volume} {111}},\ \bibinfo {pages} {034312} (\bibinfo {year} {2025})}\BibitemShut {NoStop}%
\bibitem [{\citenamefont {Lardizabal}\ and\ \citenamefont {Souza}(2015)}]{Lardizabal2015Class}%
  \BibitemOpen
  \bibfield  {author} {\bibinfo {author} {\bibfnamefont {C.~F.}\ \bibnamefont {Lardizabal}}\ and\ \bibinfo {author} {\bibfnamefont {R.~R.}\ \bibnamefont {Souza}},\ }\bibfield  {title} {\bibinfo {title} {On a class of quantum channels, open random walks and recurrence},\ }\href {https://doi.org/10.1007/s10955-015-1217-x} {\bibfield  {journal} {\bibinfo  {journal} {Journal of Statistical Physics}\ }\textbf {\bibinfo {volume} {159}},\ \bibinfo {pages} {772} (\bibinfo {year} {2015})}\BibitemShut {NoStop}%
\bibitem [{\citenamefont {Maciel}\ and\ \citenamefont {Bernardes}(2025)}]{linck25}%
  \BibitemOpen
  \bibfield  {author} {\bibinfo {author} {\bibfnamefont {P.~L.}\ \bibnamefont {Maciel}}\ and\ \bibinfo {author} {\bibfnamefont {N.~K.}\ \bibnamefont {Bernardes}},\ }\bibfield  {title} {\bibinfo {title} {Dynamics and computation in linear open quantum walks},\ }\href {https://doi.org/10.1103/lg6c-sgvn} {\bibfield  {journal} {\bibinfo  {journal} {Phys. Rev. A}\ ,\ } (\bibinfo {year} {2025})}\BibitemShut {NoStop}%
\bibitem [{\citenamefont {Konno}\ and\ \citenamefont {Yoo}(2013)}]{Konno2013Limit}%
  \BibitemOpen
  \bibfield  {author} {\bibinfo {author} {\bibfnamefont {N.}~\bibnamefont {Konno}}\ and\ \bibinfo {author} {\bibfnamefont {H.~J.}\ \bibnamefont {Yoo}},\ }\bibfield  {title} {\bibinfo {title} {Limit theorems for {Open Quantum Random Walks}},\ }\href {https://doi.org/10.1007/s10955-012-0668-6} {\bibfield  {journal} {\bibinfo  {journal} {Journal of Statistical Physics}\ }\textbf {\bibinfo {volume} {150}},\ \bibinfo {pages} {299} (\bibinfo {year} {2013})}\BibitemShut {NoStop}%
\bibitem [{\citenamefont {Attal}\ \emph {et~al.}(2015)\citenamefont {Attal}, \citenamefont {Guillotin-Plantard},\ and\ \citenamefont {Sabot}}]{Attal2015CLT}%
  \BibitemOpen
  \bibfield  {author} {\bibinfo {author} {\bibfnamefont {S.}~\bibnamefont {Attal}}, \bibinfo {author} {\bibfnamefont {N.}~\bibnamefont {Guillotin-Plantard}},\ and\ \bibinfo {author} {\bibfnamefont {C.}~\bibnamefont {Sabot}},\ }\bibfield  {title} {\bibinfo {title} {Central limit theorems for {Open Quantum Random Walks} and {Quantum Measurement Records}},\ }\href {https://doi.org/10.1007/s00023-014-0319-3} {\bibfield  {journal} {\bibinfo  {journal} {Annales Henri Poincar{\'e}}\ }\textbf {\bibinfo {volume} {16}},\ \bibinfo {pages} {15} (\bibinfo {year} {2015})}\BibitemShut {NoStop}%
\bibitem [{\citenamefont {Lloyd}\ \emph {et~al.}(2016)\citenamefont {Lloyd}, \citenamefont {Garnerone},\ and\ \citenamefont {Zanardi}}]{Lloyd16}%
  \BibitemOpen
  \bibfield  {author} {\bibinfo {author} {\bibfnamefont {S.}~\bibnamefont {Lloyd}}, \bibinfo {author} {\bibfnamefont {S.}~\bibnamefont {Garnerone}},\ and\ \bibinfo {author} {\bibfnamefont {P.}~\bibnamefont {Zanardi}},\ }\bibfield  {title} {\bibinfo {title} {Quantum algorithms for topological and geometric analysis of data},\ }\href {https://doi.org/10.1038/ncomms10138} {\bibfield  {journal} {\bibinfo  {journal} {Nat Commun}\ }\textbf {\bibinfo {volume} {7}},\ \bibinfo {pages} {10138} (\bibinfo {year} {2016})}\BibitemShut {NoStop}%
\bibitem [{\citenamefont {Schuld}\ and\ \citenamefont {Killoran}(2019)}]{Schuld2019FeatureHilbert}%
  \BibitemOpen
  \bibfield  {author} {\bibinfo {author} {\bibfnamefont {M.}~\bibnamefont {Schuld}}\ and\ \bibinfo {author} {\bibfnamefont {N.}~\bibnamefont {Killoran}},\ }\bibfield  {title} {\bibinfo {title} {Quantum machine learning in feature hilbert spaces},\ }\href {https://doi.org/10.1103/PhysRevLett.122.040504} {\bibfield  {journal} {\bibinfo  {journal} {Physical Review Letters}\ }\textbf {\bibinfo {volume} {122}},\ \bibinfo {pages} {040504} (\bibinfo {year} {2019})}\BibitemShut {NoStop}%
\bibitem [{\citenamefont {Cong}\ \emph {et~al.}(2019)\citenamefont {Cong}, \citenamefont {Choi},\ and\ \citenamefont {Lukin}}]{Cong2019QCNN}%
  \BibitemOpen
  \bibfield  {author} {\bibinfo {author} {\bibfnamefont {I.}~\bibnamefont {Cong}}, \bibinfo {author} {\bibfnamefont {S.}~\bibnamefont {Choi}},\ and\ \bibinfo {author} {\bibfnamefont {M.~D.}\ \bibnamefont {Lukin}},\ }\bibfield  {title} {\bibinfo {title} {Quantum convolutional neural networks},\ }\href {https://doi.org/10.1038/s41567-019-0648-8} {\bibfield  {journal} {\bibinfo  {journal} {Nature Physics}\ }\textbf {\bibinfo {volume} {15}},\ \bibinfo {pages} {1273} (\bibinfo {year} {2019})}\BibitemShut {NoStop}%
\bibitem [{\citenamefont {Marshall}\ \emph {et~al.}(2019{\natexlab{b}})\citenamefont {Marshall}, \citenamefont {Venuti},\ and\ \citenamefont {Zanardi}}]{Marshall2019DissipativeClassifier}%
  \BibitemOpen
  \bibfield  {author} {\bibinfo {author} {\bibfnamefont {J.}~\bibnamefont {Marshall}}, \bibinfo {author} {\bibfnamefont {L.~C.}\ \bibnamefont {Venuti}},\ and\ \bibinfo {author} {\bibfnamefont {P.}~\bibnamefont {Zanardi}},\ }\bibfield  {title} {\bibinfo {title} {Classifying quantum data by dissipation},\ }\href {https://doi.org/10.1103/PhysRevA.99.032330} {\bibfield  {journal} {\bibinfo  {journal} {Physical Review A}\ }\textbf {\bibinfo {volume} {99}},\ \bibinfo {pages} {032330} (\bibinfo {year} {2019}{\natexlab{b}})}\BibitemShut {NoStop}%
\bibitem [{\citenamefont {Korkmaz}\ \emph {et~al.}(2020)\citenamefont {Korkmaz}, \citenamefont {Türkpençe}, \citenamefont {\c{C}. Akıncı},\ and\ \citenamefont {\c{S}eker}}]{korkmaz20}%
  \BibitemOpen
  \bibfield  {author} {\bibinfo {author} {\bibfnamefont {U.}~\bibnamefont {Korkmaz}}, \bibinfo {author} {\bibfnamefont {D.}~\bibnamefont {Türkpençe}}, \bibinfo {author} {\bibfnamefont {T.}~\bibnamefont {\c{C}. Akıncı}},\ and\ \bibinfo {author} {\bibfnamefont {S.}~\bibnamefont {\c{S}eker}},\ }\bibfield  {title} {\bibinfo {title} {A thermal quantum classifier},\ }\href {https://doi.org/10.26421/QIC20.11-12-4} {\bibfield  {journal} {\bibinfo  {journal} {Quant. Inf. Comput.}\ }\textbf {\bibinfo {volume} {20}},\ \bibinfo {pages} {969} (\bibinfo {year} {2020})}\BibitemShut {NoStop}%
\bibitem [{\citenamefont {Brito}\ \emph {et~al.}(2024)\citenamefont {Brito}, \citenamefont {de~Paula~Neto},\ and\ \citenamefont {Bernardes}}]{Brito2024OQSClassifier}%
  \BibitemOpen
  \bibfield  {author} {\bibinfo {author} {\bibfnamefont {E.~B.}\ \bibnamefont {Brito}}, \bibinfo {author} {\bibfnamefont {F.~M.}\ \bibnamefont {de~Paula~Neto}},\ and\ \bibinfo {author} {\bibfnamefont {N.~K.}\ \bibnamefont {Bernardes}},\ }\bibfield  {title} {\bibinfo {title} {Quantum classifier based on open quantum systems with amplitude information loading},\ }\bibfield  {journal} {\bibinfo  {journal} {Quantum Information Processing}\ }\href {https://doi.org/10.1007/s11128-024-04526-3} {10.1007/s11128-024-04526-3} (\bibinfo {year} {2024})\BibitemShut {NoStop}%
\bibitem [{\citenamefont {Blank}\ \emph {et~al.}(2020)\citenamefont {Blank}, \citenamefont {Park}, \citenamefont {Rhee},\ and\ \citenamefont {Petruccione}}]{Blank2020TailoredKernel}%
  \BibitemOpen
  \bibfield  {author} {\bibinfo {author} {\bibfnamefont {C.}~\bibnamefont {Blank}}, \bibinfo {author} {\bibfnamefont {D.~K.}\ \bibnamefont {Park}}, \bibinfo {author} {\bibfnamefont {J.-K.~K.}\ \bibnamefont {Rhee}},\ and\ \bibinfo {author} {\bibfnamefont {F.}~\bibnamefont {Petruccione}},\ }\bibfield  {title} {\bibinfo {title} {Quantum classifier with tailored quantum kernel},\ }\href {https://doi.org/10.1038/s41534-020-0272-6} {\bibfield  {journal} {\bibinfo  {journal} {npj Quantum Information}\ }\textbf {\bibinfo {volume} {6}},\ \bibinfo {pages} {41} (\bibinfo {year} {2020})}\BibitemShut {NoStop}%
\bibitem [{\citenamefont {P{\'e}rez-Salinas}\ \emph {et~al.}(2020)\citenamefont {P{\'e}rez-Salinas}, \citenamefont {Cervera-Lierta}, \citenamefont {Gil-Fuster},\ and\ \citenamefont {Latorre}}]{PerezSalinas2020DataReuploading}%
  \BibitemOpen
  \bibfield  {author} {\bibinfo {author} {\bibfnamefont {A.}~\bibnamefont {P{\'e}rez-Salinas}}, \bibinfo {author} {\bibfnamefont {A.}~\bibnamefont {Cervera-Lierta}}, \bibinfo {author} {\bibfnamefont {E.}~\bibnamefont {Gil-Fuster}},\ and\ \bibinfo {author} {\bibfnamefont {J.~I.}\ \bibnamefont {Latorre}},\ }\bibfield  {title} {\bibinfo {title} {Data re-uploading for a universal quantum classifier},\ }\href {https://doi.org/10.22331/q-2020-02-06-226} {\bibfield  {journal} {\bibinfo  {journal} {Quantum}\ }\textbf {\bibinfo {volume} {4}},\ \bibinfo {pages} {226} (\bibinfo {year} {2020})}\BibitemShut {NoStop}%
\bibitem [{\citenamefont {Hastie}\ \emph {et~al.}(2009)\citenamefont {Hastie}, \citenamefont {Tibshirani},\ and\ \citenamefont {Friedman}}]{Hastie2009}%
  \BibitemOpen
  \bibfield  {author} {\bibinfo {author} {\bibfnamefont {T.}~\bibnamefont {Hastie}}, \bibinfo {author} {\bibfnamefont {R.}~\bibnamefont {Tibshirani}},\ and\ \bibinfo {author} {\bibfnamefont {J.}~\bibnamefont {Friedman}},\ }\href {https://doi.org/10.1007/978-0-387-84858-7} {\emph {\bibinfo {title} {The Elements of Statistical Learning: Data Mining, Inference, and Prediction}}},\ \bibinfo {edition} {2nd}\ ed.\ (\bibinfo  {publisher} {Springer},\ \bibinfo {address} {New York},\ \bibinfo {year} {2009})\BibitemShut {NoStop}%
\bibitem [{\citenamefont {Schuld}\ \emph {et~al.}(2017)\citenamefont {Schuld}, \citenamefont {Fingerhuth},\ and\ \citenamefont {Petruccione}}]{Schuld17}%
  \BibitemOpen
  \bibfield  {author} {\bibinfo {author} {\bibfnamefont {M.}~\bibnamefont {Schuld}}, \bibinfo {author} {\bibfnamefont {M.}~\bibnamefont {Fingerhuth}},\ and\ \bibinfo {author} {\bibfnamefont {F.}~\bibnamefont {Petruccione}},\ }\bibfield  {title} {\bibinfo {title} {Implementing a distance-based classifier with a quantum interference circuit},\ }\href {10.1209/0295-5075/119/60002} {\bibfield  {journal} {\bibinfo  {journal} {EPL}\ }\textbf {\bibinfo {volume} {119}},\ \bibinfo {pages} {60002} (\bibinfo {year} {2017})}\BibitemShut {NoStop}%
\bibitem [{\citenamefont {Neumann}(2021)}]{Neumann2021}%
  \BibitemOpen
  \bibfield  {author} {\bibinfo {author} {\bibfnamefont {N.~M.~P.}\ \bibnamefont {Neumann}},\ }\bibfield  {title} {\bibinfo {title} {Classification using a two-qubit quantum chip},\ }in\ \href {https://doi.org/10.1007/978-3-030-77980-1_6} {\emph {\bibinfo {booktitle} {Computational Science – ICCS 2021}}},\ \bibinfo {series} {Lecture Notes in Computer Science}, Vol.\ \bibinfo {volume} {12747},\ \bibinfo {editor} {edited by\ \bibinfo {editor} {\bibfnamefont {M.}~\bibnamefont {Paszynski}}, \bibinfo {editor} {\bibfnamefont {D.}~\bibnamefont {Kranzlmüller}}, \bibinfo {editor} {\bibfnamefont {V.~V.}\ \bibnamefont {Krzhizhanovskaya}}, \bibinfo {editor} {\bibfnamefont {J.~J.}\ \bibnamefont {Dongarra}},\ and\ \bibinfo {editor} {\bibfnamefont {P.~M.~A.}\ \bibnamefont {Sloot}}}\ (\bibinfo  {publisher} {Springer},\ \bibinfo {address} {Cham},\ \bibinfo {year} {2021})\BibitemShut {NoStop}%
\bibitem [{\citenamefont {Turkpence}\ \emph {et~al.}(2019)\citenamefont {Turkpence}, \citenamefont {\c{C}. Ak{\i}nc{\i}},\ and\ \citenamefont {\c{S}eker}}]{Turkpence2019}%
  \BibitemOpen
  \bibfield  {author} {\bibinfo {author} {\bibfnamefont {D.}~\bibnamefont {Turkpence}}, \bibinfo {author} {\bibfnamefont {T.}~\bibnamefont {\c{C}. Ak{\i}nc{\i}}},\ and\ \bibinfo {author} {\bibfnamefont {S.}~\bibnamefont {\c{S}eker}},\ }\bibfield  {title} {\bibinfo {title} {A steady state quantum classifier},\ }\href {https://doi.org/10.1016/j.physleta.2019.02.018} {\bibfield  {journal} {\bibinfo  {journal} {Physics Letters A}\ }\textbf {\bibinfo {volume} {383}},\ \bibinfo {pages} {1410} (\bibinfo {year} {2019})}\BibitemShut {NoStop}%
\bibitem [{\citenamefont {Zhang}\ \emph {et~al.}(2021)\citenamefont {Zhang} \emph {et~al.}}]{Zhang2021}%
  \BibitemOpen
  \bibfield  {author} {\bibinfo {author} {\bibfnamefont {J.}~\bibnamefont {Zhang}} \emph {et~al.},\ }\bibfield  {title} {\bibinfo {title} {Interactive quantum classifier inspired by quantum open system theory},\ }in\ \href {https://doi.org/10.1109/IJCNN52387.2021.9534374} {\emph {\bibinfo {booktitle} {2021 International Joint Conference on Neural Networks (IJCNN)}}}\ (\bibinfo {year} {2021})\ pp.\ \bibinfo {pages} {1--7}\BibitemShut {NoStop}%
\bibitem [{\citenamefont {Venegas-Andraca}(2012)}]{Venegas12}%
  \BibitemOpen
  \bibfield  {author} {\bibinfo {author} {\bibfnamefont {S.~E.}\ \bibnamefont {Venegas-Andraca}},\ }\bibfield  {title} {\bibinfo {title} {Quantum walks: a comprehensive review},\ }\href {https://doi.org/10.1007/s11128-012-0432-5} {\bibfield  {journal} {\bibinfo  {journal} {Quantum Inf Process}\ }\textbf {\bibinfo {volume} {11}},\ \bibinfo {pages} {1015–} (\bibinfo {year} {2012})}\BibitemShut {NoStop}%
\bibitem [{\citenamefont {Aharonov}\ \emph {et~al.}(1993)\citenamefont {Aharonov}, \citenamefont {Davidovich},\ and\ \citenamefont {Zagury}}]{Ahar93}%
  \BibitemOpen
  \bibfield  {author} {\bibinfo {author} {\bibfnamefont {Y.}~\bibnamefont {Aharonov}}, \bibinfo {author} {\bibfnamefont {L.}~\bibnamefont {Davidovich}},\ and\ \bibinfo {author} {\bibfnamefont {N.}~\bibnamefont {Zagury}},\ }\bibfield  {title} {\bibinfo {title} {Quantum random walks},\ }\href {https://doi.org/10.1103/PhysRevA.48.1687} {\bibfield  {journal} {\bibinfo  {journal} {Phys. Rev. A}\ }\textbf {\bibinfo {volume} {48}},\ \bibinfo {pages} {1687} (\bibinfo {year} {1993})}\BibitemShut {NoStop}%
\bibitem [{\citenamefont {Portugal}(2018)}]{QWSA}%
  \BibitemOpen
  \bibfield  {author} {\bibinfo {author} {\bibfnamefont {R.}~\bibnamefont {Portugal}},\ }\href@noop {} {\emph {\bibinfo {title} {Quantum Walks and Search Algorithms}}},\ \bibinfo {edition} {2nd}\ ed.\ (\bibinfo  {publisher} {Springer},\ \bibinfo {year} {2018})\BibitemShut {NoStop}%
\bibitem [{\citenamefont {Childs}(2009)}]{Childs09}%
  \BibitemOpen
  \bibfield  {author} {\bibinfo {author} {\bibfnamefont {A.~M.}\ \bibnamefont {Childs}},\ }\bibfield  {title} {\bibinfo {title} {Universal computation by quantum walk},\ }\href {https://doi.org/10.1103/PhysRevLett.102.180501} {\bibfield  {journal} {\bibinfo  {journal} {Phys. Rev. Lett.}\ }\textbf {\bibinfo {volume} {102}},\ \bibinfo {pages} {180501} (\bibinfo {year} {2009})}\BibitemShut {NoStop}%
\bibitem [{\citenamefont {Lovett}\ \emph {et~al.}(2010)\citenamefont {Lovett}, \citenamefont {Cooper}, \citenamefont {Everitt}, \citenamefont {Trevers},\ and\ \citenamefont {Kendon}}]{Neil10}%
  \BibitemOpen
  \bibfield  {author} {\bibinfo {author} {\bibfnamefont {N.~B.}\ \bibnamefont {Lovett}}, \bibinfo {author} {\bibfnamefont {S.}~\bibnamefont {Cooper}}, \bibinfo {author} {\bibfnamefont {M.}~\bibnamefont {Everitt}}, \bibinfo {author} {\bibfnamefont {M.}~\bibnamefont {Trevers}},\ and\ \bibinfo {author} {\bibfnamefont {V.}~\bibnamefont {Kendon}},\ }\bibfield  {title} {\bibinfo {title} {Universal quantum computation using the discrete-time quantum walk},\ }\href {https://doi.org/10.1103/PhysRevA.81.042330} {\bibfield  {journal} {\bibinfo  {journal} {Phys. Rev. A}\ }\textbf {\bibinfo {volume} {81}},\ \bibinfo {pages} {042330} (\bibinfo {year} {2010})}\BibitemShut {NoStop}%
\bibitem [{\citenamefont {Breuer}\ and\ \citenamefont {Petruccione}(2010)}]{OQS}%
  \BibitemOpen
  \bibfield  {author} {\bibinfo {author} {\bibfnamefont {H.}~\bibnamefont {Breuer}}\ and\ \bibinfo {author} {\bibfnamefont {F.}~\bibnamefont {Petruccione}},\ }\href@noop {} {\emph {\bibinfo {title} {The Theory of Open Quantum Systems}}}\ (\bibinfo  {publisher} {Oxford Academic},\ \bibinfo {year} {2010})\BibitemShut {NoStop}%
\bibitem [{\citenamefont {Behrends}(2000)}]{MC}%
  \BibitemOpen
  \bibfield  {author} {\bibinfo {author} {\bibfnamefont {E.}~\bibnamefont {Behrends}},\ }\href@noop {} {\emph {\bibinfo {title} {Introduction to Markov Chains}}}\ (\bibinfo  {publisher} {Springer},\ \bibinfo {year} {2000})\BibitemShut {NoStop}%
\bibitem [{\citenamefont {Kampen}(2007)}]{SP}%
  \BibitemOpen
  \bibfield  {author} {\bibinfo {author} {\bibfnamefont {N.~G.~V.}\ \bibnamefont {Kampen}},\ }\href@noop {} {\emph {\bibinfo {title} {Stochastic Processes in Physics and Chemistry}}},\ \bibinfo {edition} {3rd}\ ed.\ (\bibinfo  {publisher} {North-Holland},\ \bibinfo {year} {2007})\BibitemShut {NoStop}%
\bibitem [{\citenamefont {Gardiner}(1985)}]{HSM}%
  \BibitemOpen
  \bibfield  {author} {\bibinfo {author} {\bibfnamefont {C.~W.}\ \bibnamefont {Gardiner}},\ }\href@noop {} {\emph {\bibinfo {title} {Handbook of Stochastic Methods}}},\ \bibinfo {edition} {2nd}\ ed.\ (\bibinfo  {publisher} {Springer},\ \bibinfo {year} {1985})\BibitemShut {NoStop}%
\end{thebibliography}%

\end{document}